\documentstyle[graphicx,natbib,epsfig,amsmath,amssymb,traditabstract]{aa} 

\newcommand{\micron}{\hbox{$\mu$m}}

\newcommand{\tkin}{\,\hbox{$T_{\rm kin}$}}

\newcommand{\msun}{\,\hbox{$M_{\odot}$}}
\newcommand{\lsun}{\,\hbox{$L_{\odot}$}}
\newcommand{\kms}{\,\hbox{\hbox{km}\,\hbox{s}$^{-1}$}}

\newcommand{\htwo}{\,\hbox{$\rm{H_ 2}$}}

\newcommand{\spi}{{\it Spitzer}}
\newcommand{\ang}{\,\hbox{\AA}}

\newcommand{\um}{\,\hbox{$\mu$m}}

\newcommand{\cooz}{\hbox{CO(1-0)}}
\newcommand{\coto}{\hbox{CO(2-1)}}

\newcommand{\cott}{\hbox{CO(3-2)}}
\newcommand{\her}{\hbox{\it Herschel}}

\begin{document}
\textheight=24.8cm
\addtolength{\topmargin}{-.2cm}
\setlength{\parskip}{0.5mm plus 0.0mm minus0.0mm}
  
   \title{Heating of the molecular gas in the massive outflow of the local ultraluminous-infrared and radio-loud galaxy 4C12.50}

   \subtitle{}

   \author{K. M. Dasyra\inst{1}, F. Combes\inst{1}, G. S. Novak\inst{1},  M. Bremer\inst{2}, L. Spinoglio\inst{3}, M. Pereira Santaella\inst{3},  P. Salom\'e\inst{1},  \and  E. Falgarone\inst{1}
	  }

   \institute{
             Observatoire de Paris, LERMA (UMR8112), 61 Av. de l'Observatoire, F-75014, Paris, France 
             \and
             Institut de Radio Astronomie Millim\'etrique, 300 rue de la Piscine, F-38406 Saint-Martin d'H\`eres, France 
             \and
             Istituto di Astrofisica e Planetologia Spaziali, INAF-IAPS, Via Fosso del Cavaliere 100, I-00133 Roma, Italy \\
             }
   \date{}
   
    \abstract 
    {  
     We present a comparison of the molecular gas properties in the outflow vs. in the ambient medium of the local prototype radio-loud 
     and ultraluminous-infrared galaxy 4C12.50 (IRAS13451$+$1232), using new data from the IRAM Plateau de Bure interferometer 
     and 30\,m telescope, and the \her\ space telescope.  Previous \htwo\ (0-0) S(1) and S(2) observations with the \spi\ space telescope 
     had indicated that the warm ($\sim$400\,K) molecular gas in 4C12.50 is made up of a 1.4($\pm$0.2)$\times$10$^8$\msun\  ambient 
     reservoir and a 5.2($\pm$1.7)$\times$10$^7$\msun\ outflow. The new \cooz\ data cube indicates that the corresponding cold (25\,K) 
     \htwo\ gas mass is 1.0($\pm$0.1)$\times$10$^{10}$\msun\ for the ambient medium and $<$1.3$\times$10$^8$\msun\ for the outflow, 
     when using a CO-intensity-to-H$_2$-mass conversion factor $\alpha$ of 0.8\msun /(K\kms\,pc$^2$).  The combined mass outflow rate 
     is high, 230-800\msun /yr, but the amount of gas that could escape the galaxy is low. A potential inflow of gas from a 
     3.3($\pm$0.3)$\times$10$^8$\msun\ tidal tail could moderate any mass loss. The mass ratio of warm-to-cold molecular gas is 
     $\gtrsim$30 times higher in the outflow than in the ambient medium, indicating that a non-negligible fraction of the accelerated gas is 
     heated to temperatures at which star formation is inefficient. This conclusion is robust against the use of different $\alpha$ factor values, 
     and/or different warm gas tracers (\htwo\ vs. \htwo\ plus CO): with the CO-probed gas mass being at least 40 times lower at 400\,K 
     than at 25\,K, the total warm-to-cold mass ratio is always lower in the ambient gas than in the entrained gas. Heating of the molecular gas 
     could facilitate the detection of new outflows in distant galaxies by enhancing their emission in intermediate rotational number CO lines.
     }
        \keywords{  ISM: jets and outflows ---
   			ISM: kinematics and dynamics ---
   			Line: profiles ---
   			Galaxies: active ---
   			Galaxies: nuclei ---
   			Infrared: galaxies
             		 }

   \titlerunning{ Heating of the molecular gas in the massive outflow of 4C12.50}
   \authorrunning{ Dasyra et al.}
   
   \maketitle


\section{Introduction}
\label{sec:intro}

Mass accretion events onto black holes (BHs) can release more energy than the binding energy of gas in galaxies \citep{fabian12},
and part of this energy can be deposited onto the interstellar medium (ISM) through radiation, accretion-disk winds, and radio jets. 
The energy carried by these so-called feedback mechanisms can excite or kinematically distort gas on large scales \citep[e.g., over 
areas of square kiloparsecs;][]{lipari04,fu09,rupke11,westmoquette12}. It can even affect galaxies external 
to the active galactic nucleus (AGN) that generated it \citep[e.g.,][]{croft06,cantalupo12}. The effects of feedback on the evolution of 
galaxies are most severe if part of the gas is accelerated beyond escape velocity. The high-velocity gas will then be lost to the intergalactic 
medium. Any star formation associated with it will be quenched. For the remaining gas, star formation will be delayed. The 
low-velocity gas will cool radiatively, resettle in a disk, and restart its collapse towards the formation of new cores. For these reasons, AGN 
feedback has been considered as a source of negative feedback, capable of regulating galaxy growth. Positive feedback, i.e., compression 
of gas leading to an enhancement of star formation has also been occasionally suggested \citep[e.g.,][]{vanbreugel85,gaibler12,silk13}. 

AGN feedback,  implemented either as a source of heating coupled to the hot gas or as a source of momentum coupled to the cold gas, 
was included in cosmological simulations to suppress star formation at earlier epochs and to make the predicted number counts of massive 
present-day galaxies match the observed ones \citep{granato04,croton06,bower06,somerville08,booth09,debuhr12}.  The claim that 
feedback could change the high-mass end of the local galaxy mass function by more than an order of magnitude was however contested 
by \citet{birnboim07}, \citet{dekel08}, and \citet{bernardi13}, who argued that the numbers could be revised downwards through, e.g., 
different stellar population modeling assumptions. 

\begin{figure}[t!]
\begin{center}
\includegraphics[width=\columnwidth]{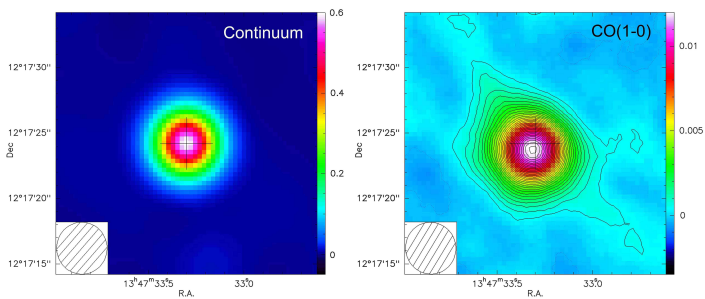} 
\caption{ 
{\it Left:} 102 GHz continuum image of 4C12.50 (in units of Jy/beam). {\it Right:} Continuum-free \cooz\ line intensity map of 4C12.50 averaged over the -500 \kms\ to 500 \kms\ range, 
showing extended emission that is offset from the continuum peak (marked with a cross). The contours are at the 3$\sigma$ levels for this image, i.e., at steps of 0.36\,mJy/beam.}
\label{fig:CO_cont_line} 
\end{center}
\end{figure}

The change in the galaxy mass function by AGN feedback remains to be quantified or constrained by observations. 
To quantify it, we need to determine the frequency of AGN-driven outflows that can considerably delay or quench 
star formation in their host galaxies. Even though high-velocity outflows have long been identified in optical and radio data of ionized and 
neutral atomic gas tracers in local AGN \citep[e.g.,][]{heckman00,rupke05,martin05,morganti05,holt08}, this research field became particularly 
appealing thanks to the launch of infrared (IR) space missions (e.g., \her ) and the improvement on the sensitivity and the spectral range of 
ground-based mm facilities, which enabled us to probe the molecular gas content of outflows.  Being the densest ISM component, the molecular 
gas can considerably contribute to the mass of the entrained gas, if not dominate it.  Motions of molecular gas that could be associated with a nuclear
outflow have been presented for $\sim$50 local galaxies to date \citep{sakamoto06,sakamoto09,leon07,feruglio10,fischer10,alatalo11,rangwala11,
sturm11,krips11,dasyra_combes11,dasyra_combes12, aalto12,tsai12,morganti13a,combes13,spoon13,veilleux13,cicone14}. Mass measurements 
of the accelerated gas were presented in $\sim$one fifth of the cases and found to be in the range 10$^6$-10$^9$\msun .

To identify spectral lines with strong outflow signatures, and use them for the systematic and efficient detection of new outflows, we need to 
perform excitation studies of the gas in the flow, motivated by the argument that the AGN feedback could leave both kinematic and excitation signatures 
on the ISM. Such studies have nonetheless only been presented for a couple of sources.  In Mrk 231, \citet{cicone12} found no evidence for a strong 
contribution of shocks on the excitation of outflowing CO molecules at states of low rotational number $J$: the \coto /\cooz\  flux ratio was comparable in 
the outflow and in the ambient medium. 

In this paper, we present evidence of heating of the molecular gas in the outflow of the local radio-loud and ultraluminous-infrared system 4C12.50 
(IRAS 13451$+$1232), i.e., a late-stage merger of two galaxies where the nuclei are separated by 4.4\,kpc and surrounded by smaller structures
\citep{batcheldor07}.  A relativistic radio jet with an intrinsic bulk flow speed of at least 0.8$c$ emerges from the system's west nucleus \citep{lister03}. 
The setup of a radio jet propagating through a gas-rich  galaxy, which is rare in the local Universe but common at high redshift \citep[e.g.,][]{sajina08,ivison12}, 
makes 4C12.50 an excellent target for studying the properties of the outflowing molecular gas. The uniqueness of 4C12.50 lies in the discovery and the
characterization of a warm molecular gas outflow in it: it was the only source in the \spi\ archive for which an outflow-related wing was detected 
in two purely rotational \htwo\ lines, providing the entrained gas excitation temperature and mass \citep{dasyra_combes11}. Cold gas was also discovered in 
the outflow thanks to absorption seen in CO(2-3) with a significant optical deph against the millimeter continuum source \citep{dasyra_combes12}. In this paper, 
we compare the properties of the gas in the outflow vs. that in the ambient medium using new submillimeter and millimeter data, and we discuss the implications of 
our findings for future searches of massive molecular gas outflows. Throughout our work, we adopt a $\Lambda$CDM cosmology with H$_0$=70 \kms\  Mpc$^{-1}$, 
$\Omega_{M}$=0.3, and $\Omega_{\Lambda}$=0.7.

\begin{figure}
\begin{center}
\includegraphics[width=8cm,height=6.7cm]{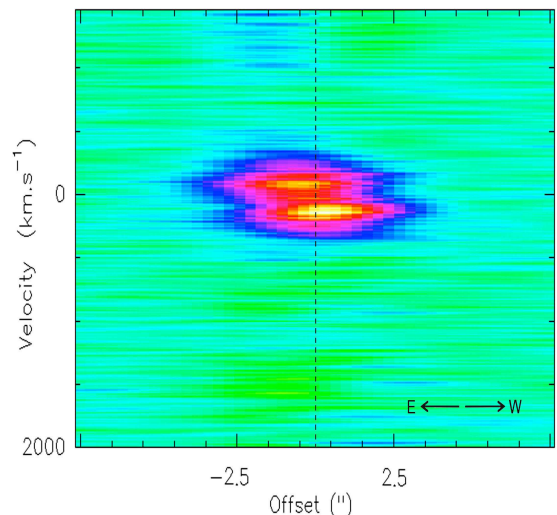} 
\caption{Position-velocity diagram of the continuum-free \cooz\ emission along the east-west axis, displayed for a velocity resolution of 6\kms . 
The x-axis offset is computed from the radio core location, which is marked with a dashed line. The offset increases (/decreases) towards the west (/east).}
\label{fig:pos_vel}
\end{center}
\end{figure}
\begin{figure*}
\begin{center}
\includegraphics[width=19.3cm]{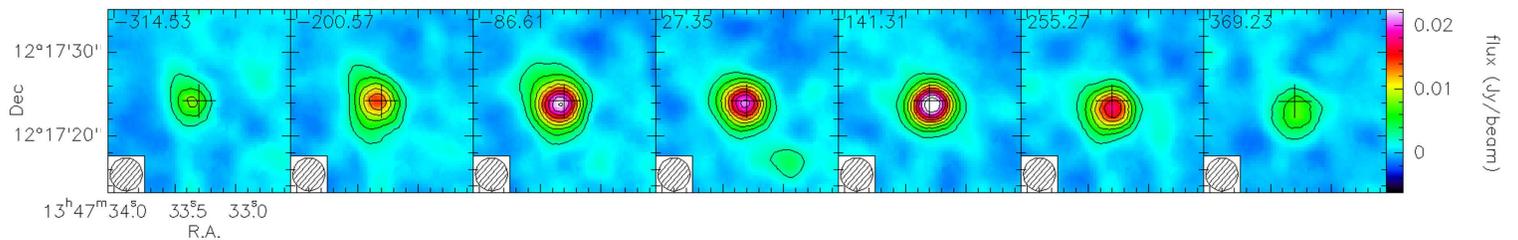} 
\caption{Continuum-free \cooz\ line intensity map of 4C12.50, displayed for $\sim$114 \kms\ bins (in the range $-$315\kms\ to 369\kms ). The number at 
the upper-left corner of each frame corresponds to the bin mean velocity. Contours are plotted at integer multiples of the 5$\sigma$ level for these images, 
i.e., at steps of 2.4 mJy/beam. A shift towards the west of the line peak position from negative to positive velocities, as well as a distinct kinematic component 
near rest-frame velocity are observed.}
\label{fig:COmap_Vshift}
\end{center}
\end{figure*}
\begin{figure*}
\begin{center}
\includegraphics[width=19.3cm]{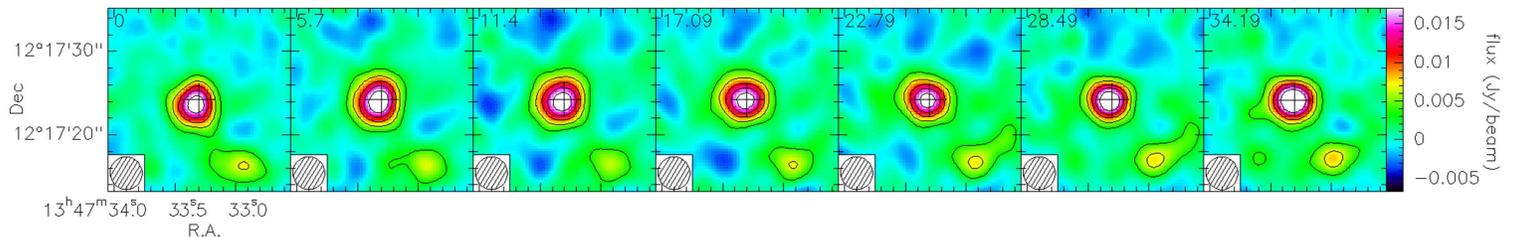} 
\caption{ Same as in Fig.~\ref{fig:COmap_Vshift}, but for bins of 17\kms\ (in the range 0 to 34\kms ). Contours are plotted at integer multiples of the 3.5$\sigma$ 
level for these images, i.e., at steps of 3.3 mJy/beam.}
\label{fig:COmap_structure}
\end{center}
\end{figure*}

\section{Data}
\label{sec:data}

The \cooz\ observations were carried out with the Plateau de Bure (PdB) Interferometer of the Institut de Radioastronomie 
Millim\'etrique (IRAM) as part of the program V028 (PI Dasyra). The data were taken on four different days between 
September 2011 and January 2012, during which four, five, or six antennae were placed in compact (C, D, or 
special) configurations. The dual polarization SIS receivers were tuned at a frequency $\nu$ of 102.761\,GHz, 
and used with the WIDEX backend. The typical weather conditions were 0.6-2.2\arcsec\ of seeing and 2-10\,mm of water vapor. 
Bandpass calibrators (0923$+$392,1354$+$195, 3C84, 3C279) were observed before our primary target during the first two 
runs. During the next two runs, they were observed before and after each 22.5 minute$-$long integration on 4C12.50. This 
change to the observational strategy enabled us to minimize the scatter due to the strong millimeter continuum of 4C12.50 by 
monitoring changes in the time$-$variable bandpass solution and thus increasing the accuracy of the subsequent phase and 
flux calibrations. To perform the latter two tasks, we used the default IRAM environment, CLIC\footnote{\label{footnoteIRAM}
http://www.iram.fr/IRAMFR/GILDAS}, selecting 4C12.50 itself for the phase calibration and MWC\,349 for the flux calibration. 
The calibrated visibilities of all tracks were merged into a single table, of depth equivalent to 12.5\,hours of observations with a 
six-antenna array. A cube of 0.34\arcsec\ per spatial pixel was created from this table using the default PdB image reconstruction 
environment, MAPPING$^{\ref{footnoteIRAM}}$.  Its 1$\sigma $ noise is 1.1 mJy per beam at a spectral resolution of 5.7\kms . The 
clean 3\,mm beam corresponds to an ellipse with semi$-$major and semi$-$minor axis of 4.0\arcsec\ and 3.8\arcsec , respectively, 
at a position angle of 14$^{\circ} $. We constructed the continuum image by computing the median of all cube planes in the 450 \kms\ 
to 1250 \kms\ range, and removed it from the entire cube to obtain a continuum-free \cooz\ line cube (Figure~\ref{fig:CO_cont_line}).

Observations at 205.522\,GHz targeting the \coto\ line were carried out in June 2012 with the IRAM 30\,m telescope. The Eight 
Mixer Receiver \citep[EMIR;][]{carter12} and the Wideband Line Multiple Autocorrelator (WILMA) backend were used for this program.
The FTS backend was also connected for repeatability. The data presented here are rather shallow, of 20 minutes of on-source integration 
time, taken while the 1.5\,mm system temperature varied from 400 to 450\,K. The deterioration of the weather conditions prevented the 
acquisition of more data. The individual scans were examined, stacked, and binned to 47\kms\ channels within the CLASS environment$^{\ref{footnoteIRAM}}$.  

We followed up 4C12.50 with the PACS \citep{poglitsch10} far-infrared instrument on board \her\ \citep{pilbratt10} to look for emission from CO molecules that are excited to states of high rotational number. High spectral sampling CO(16-15) observations were carried out for our GT2\_lspinogl\_6 program in range-spectroscopy mode using chop-nod cycles. To populate the CO spectral line energy distribution (SLED), we combined our data with all PACS range or line spectroscopy data in the \her\ archive that cover the observed-frame wavelengths of CO lines in 4C12.50. For this reason, we examined the blue array data and their parallel red array data in each astronomical observation request (AOR). We found that, in total, information is available for four transitions, CO(33-32), CO(22-21), CO(18-17), and CO(16-15), with integration times of 1.2, 4.0, 1.4, and 2.1 hours, respectively. For CO(22-21), this is the total time of the combined OT1\_dfarrah\_1 and OT1\_sveilleu\_3 program data. We reduced the data within the \her\ Interactive Processing Environment (HIPE; version 8.3, calibration tree version 32; \citealt{ott10}), by running the default PACS spectrometer pipeline for chopped-line and short-range scans with a wavelength grid oversampling parameter\footnote{The differences between the spectra obtained by running the default pipeline and the spectra obtained by running the routines for an alternative flux calibration based on the telescope background are within the given error bars. } of 2. We flagged as outliers all pixel 
read-outs in emission-line-free wavelengths whose flux exceeded 5$\sigma$. Because 4C12.50 is a point source for PACS, we obtained its final spectrum at each
wavelength by extracting the spectrum of the central 9.4"$\times$9.4" spaxel of each data cube, and by multiplying it by the point-source-correction factor in order to recover its full flux. We followed an identical procedure to reduce all other archival PACS observations of 4C12.50 in order to obtain information on its far-infrared dust continuum emission. From this process, we excluded AORS in spectral ranges that suffer from flux leakage from/to other wavelengths, such as $<$54\um , 70$-$74\um , and 95-105\um .

\begin{figure}[ht!]
\begin{center}
\includegraphics[width=\columnwidth]{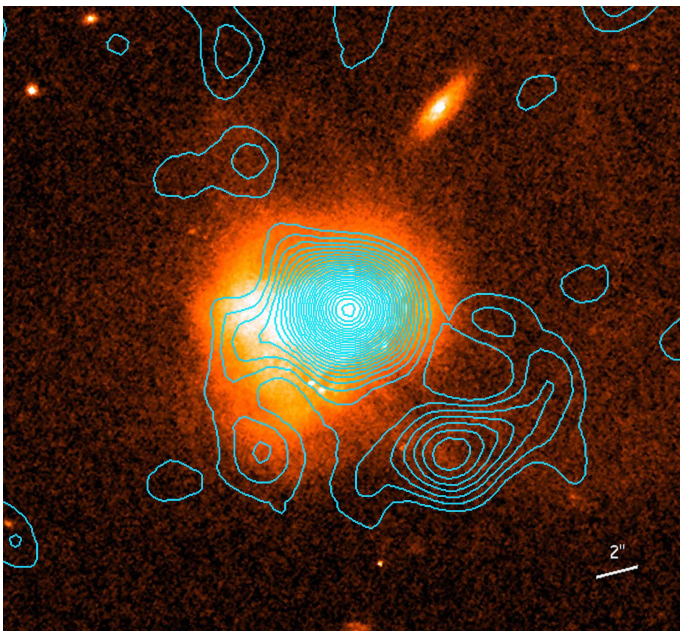} 
\caption{ Contours of the \cooz\ emission of 4C12.50, integrated from 25\kms\ to 35\kms\ and plotted over an 7300\ang\ {\it Hubble} Space Telescope ACS image 
\citep{batcheldor07}. The contours start at the 1\,$\sigma$ level of the collapsed \cooz\ image, 0.014 Jy/beam. They are spaced by an equal amount. }
\label{fig:hst_pdb_contours}
\end{center}
\end{figure}


\section{Results}
\label{sec:results}

\subsection{CO(1-0) data: gas mass and distribution}

The median channel with a \cooz\ detection in the PdB data cube is at 102.756\,GHz, pointing to a redshift $z$ of 0.12179. When adopting 
this redshift and collapsing the continuum-subtracted cube over the $-$500\kms\ to 500\kms\ velocity range, we find that the \cooz\ emission 
is offset from the 3\,mm continuum peak and marginally extended (Figure~\ref{fig:CO_cont_line}). When examining the data cube within the 
$-$500$<$V$<$500\kms\ range, we find that the \cooz\ emission progresses from the east to the west as the velocity increases from negative 
to positive values (Figures~\ref{fig:pos_vel},\ref{fig:COmap_Vshift}). It peaks at two locations, $\sim$1\arcsec\ (or $\sim$2.2 kpc) apart. This 
emission arises from gas that could be associated with the two main merging nuclei of 4C12.50 \citep{dasyra06a}, with an off-center disk forming 
due to the merger, or with a combination of the above. 

A distinct, off-center component is also detected in the central velocity bin of Figure~\ref{fig:COmap_Vshift}, at $\sim$10\,kpc south-west of the main 
nucleus. An inspection of the data cube at a higher spectral resolution (Figure~\ref{fig:COmap_structure}) reveals that this component has a tidal structure, 
approaching the main component at systemic velocity. Its distance from the center progressively increases with increasing velocity, until it reaches 
12 kpc at $\sim$30\kms . The structure has a faint counterpart in the optical (Figure~\ref{fig:hst_pdb_contours}), and it is by far the most gas rich 
(and thus dust rich) of all tidal tails in this system.

\begin{figure*}[t]
\begin{center}
\includegraphics[width=15.0cm]{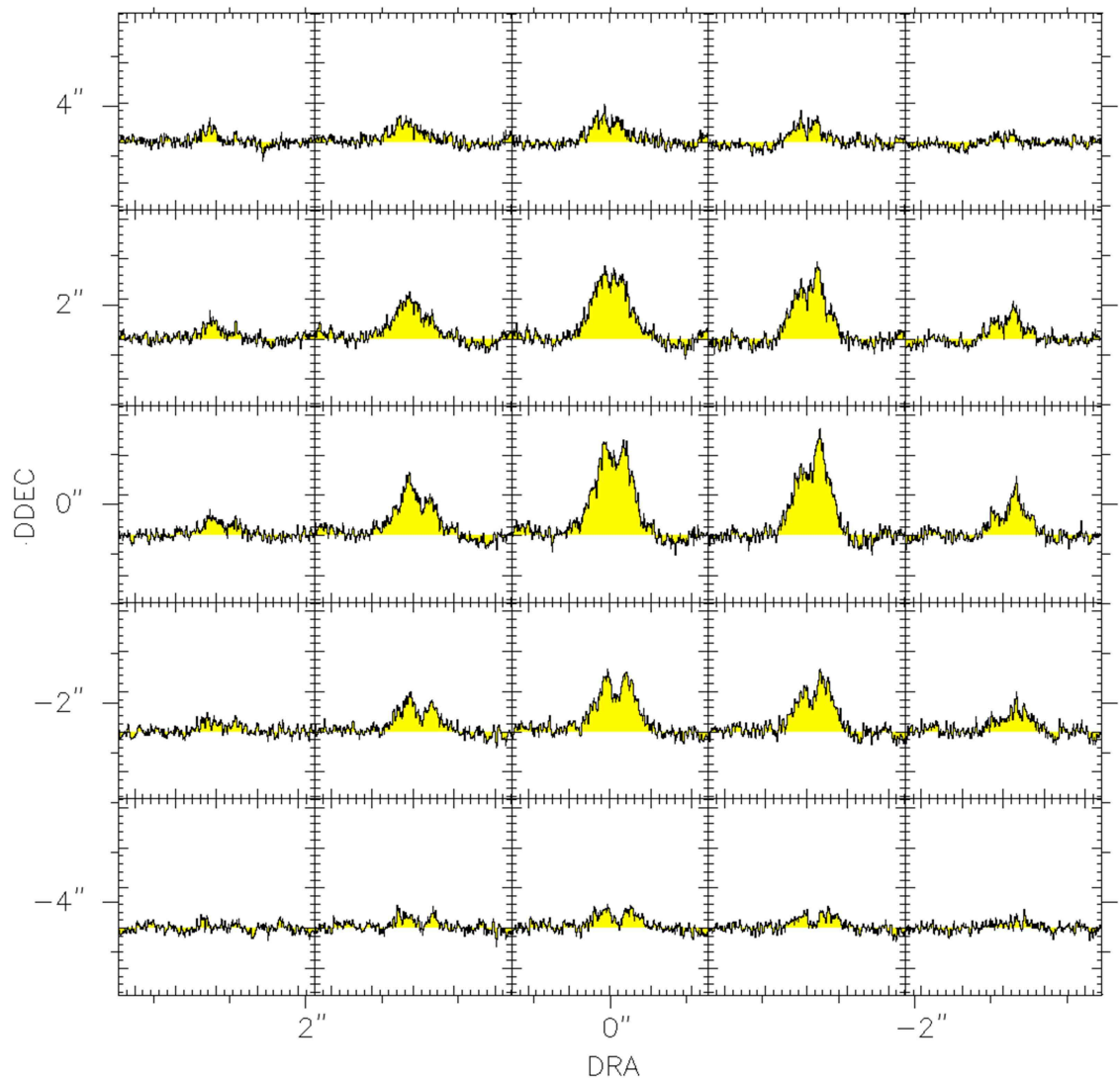} 
\caption{ \cooz\ line profile changes within few arcseconds (i.e., at sub-beam scales) from the position of the radio core in the PdB data. All spectra are plotted in the $-$1200\kms\ to 1200\kms\ velocity range.}
\label{fig:CO10_main_kinematics}
\end{center}
\end{figure*}

\begin{figure}[ht!]
\begin{center}
\includegraphics[width=8cm]{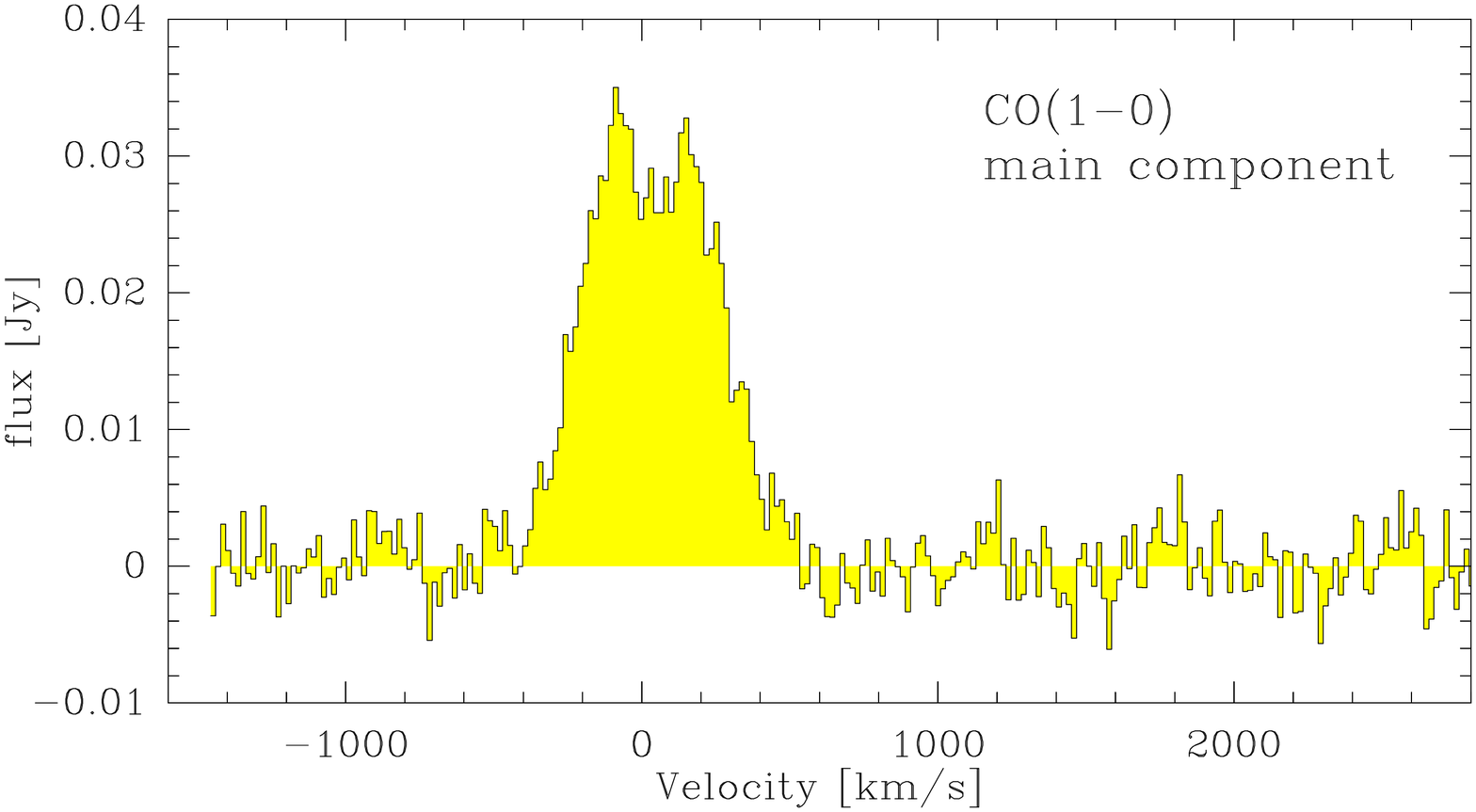} 
\caption{ Line profile of the main CO(1-0) emission component in the PdB data. The flux is integrated down to the 1$\sigma$ level of 
the integrated \cooz\ emission (Figure~\ref{fig:CO_cont_line}; typically 7-9\arcsec\ from the radio core), excluding the emission from the tidal 
structure, which is shown in Figure~\ref{fig:CO10fluxtail}.
}
\label{fig:CO10flux}
\end{center}
\end{figure}

\begin{figure}[h!]
\begin{center}
\includegraphics[width=8cm]{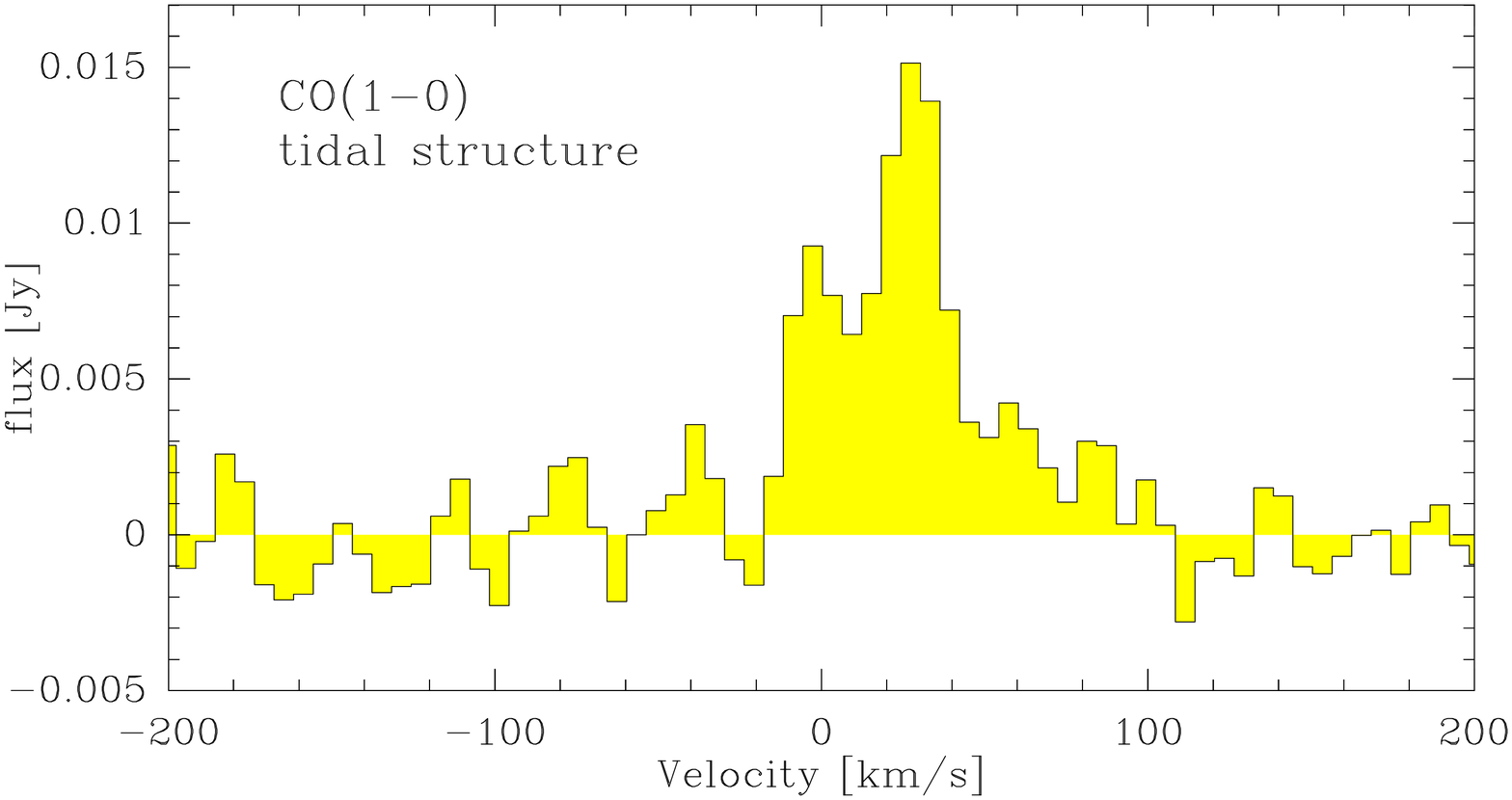} 
\caption{  PdB \cooz\ spectrum of the tidal structure south-west of the merging nuclei of 4C12.50, integrated down to the 1$\sigma$ flux level of Figure~\ref{fig:hst_pdb_contours}. 
}
\label{fig:CO10fluxtail}
\end{center}
\end{figure}

The spectra extracted from the data cube are shown in Figures~\ref{fig:CO10_main_kinematics} and \ref{fig:CO10flux} for the main component, and 
Figure~\ref{fig:CO10fluxtail} for the tidal tail. Figure~\ref{fig:CO10_main_kinematics} reveals an asymmetry in the \cooz\ emission profile along the
east-west axis at sub-beam scales that is likely due to the merger. The profile can also be compatible with absorption of CO and continuum emission 
by molecules at systemic velocity, as, e.g., in Centaurus A \citep{eckart90,wiklind97}. 

To find the mass of the \htwo\ gas that is associated with the CO emission shown in Figures~\ref{fig:CO10flux} and~\ref{fig:CO10fluxtail}, we computed the product 
\hbox{3.25$\times$10$^7$\,$\alpha$\,$S_{\rm {CO(1-0)}}$\,$\Delta V$\,$\nu_{\rm obs}^{-2}$\,{$D_L^2$} /{(1+$z$)$^3$}} \msun ,  where $\alpha$ is the 
CO intensity to \htwo\ mass conversion factor, $S_{\rm {CO(1-0)}}$\,$\Delta V$ is the integrated line flux in Jy\kms , $\nu_{\rm obs}$ is the observed \cooz\ 
frequency in GHz, and $D_L$ is the source luminosity distance in Mpc \citep{solomon97}. We adopted an $\alpha$ value of 0.8\msun /(K\kms\,pc$^2$), to 
be consistent with the literature assuming a lower conversion factor for ultraluminous infrared galaxies (ULIRGs) than for the Milky Way \citep{downes98}. 

For the main emission component (Figures~\ref{fig:CO_cont_line},~\ref{fig:CO10flux}), we measure a flux of 18.2$\pm$1.5\,Jy\kms . From this flux we 
calculate a cold \htwo\ gas mass reservoir of 1.0\,($\pm$0.1)$ \times$10$^{10}$\msun . Likewise, the flux in the tidal tail is 0.61$\pm$0.04 Jy\kms\ 
(integrated down to the 1$\sigma$ level, as shown in Figure~\ref{fig:hst_pdb_contours}). This 
corresponds to an \htwo\ mass of 3.3\,($\pm$0.3)$ \times$10$^{8}$\msun . We thus find that a mass comparable to 1/30$^{\rm th}$ of the main reservoir 
could be infalling thanks to the tidal tail (see Section~\ref{sec:discussion}). Because the total \cooz\ flux in the PdB data agrees within the error with that 
previously measured in IRAM 30\,m telescope data \citep[20$\pm$6\,Jy\kms ;][]{dasyra_combes11}, we deduce that the interferometer did not miss a 
significant part of the extended emission.

A spatially-unresolved absorption line at $-$720\kms\ at the radio core and less than a handful of emission lines at V$<$$-$500\kms\ or V$>$500\kms\ 
were seen in the PdB data at $\sim$3-5 root mean square noise ($\sigma$) levels. Their widths never exceeded $\sim$80\kms . These lines could be 
real (due to a small collection of high-velocity clouds in the outflow), or artifacts (due to the varying position and intensity of the mm continuum, 
which leads to an inadequate removal of the point spread function). A reliable detection of the outflow in \cooz\ remains to be achieved with 
higher dynamic range observations. In the rest of this work, we adopt an upper limit of 1.3$\times$10$^{8}$\msun\  for the mass of the cold molecular 
gas in the outflow. We computed this limit by integrating 3$\sigma$ over the velocity range that the molecular outflow spans in other 
wavelengths ($-$1300$\kms <$V$<$$-$300\kms ;  \citealt{dasyra_combes11,dasyra_combes12}). We further assumed that the outflow is spatially 
unresolved, constrained within our four-kiloparsec-radius beam. Indeed, no evidence of an extended (ionized gas) outflow was found in multi-slit optical 
spectra of 4C12.50 \citep{holt08,holt11}. The neutral atomic gas component of the outflow observed in the radio is found $\sim$150\,pc away from 
the nucleus, at the tip of the southern radio hot spot \citep{morganti13b}. 

\subsection{Higher rotational number CO lines}

\begin{figure}[h!]
\begin{center}
\includegraphics[width=8cm]{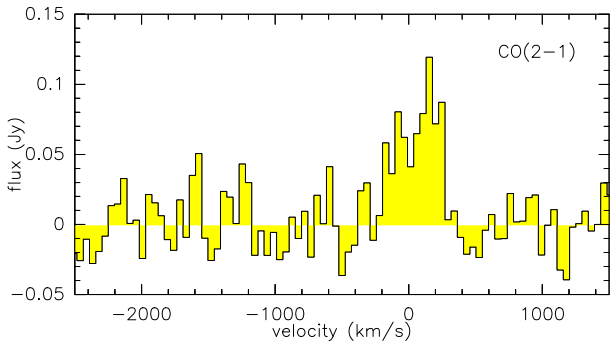} 
\caption{ IRAM 30\,m telescope data of \coto, comprising all regions where \cooz\ was detected in the PdB data (i.e., the main component and
the tidal structure).} 
\label{fig:CO21}
\end{center}
\end{figure}

\begin{figure*}
\begin{center}
\includegraphics[width=16.3cm]{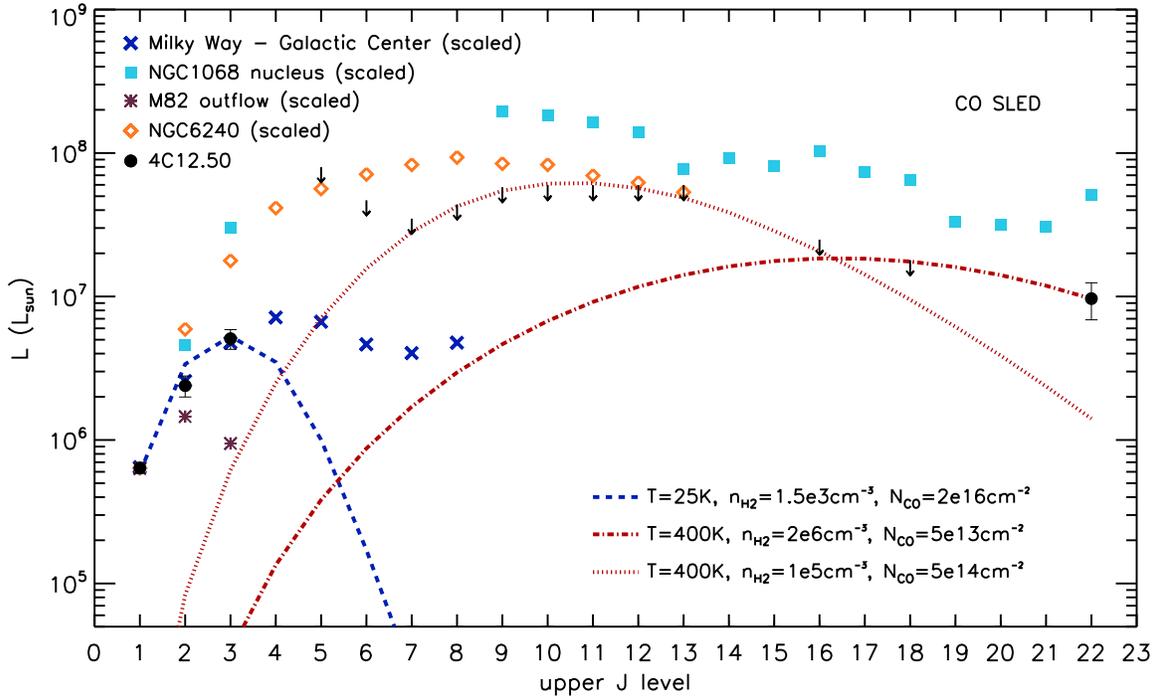} 
\caption{ CO SLED of 4C12.50 shown for transitions up to $J$=22-21. 
The CO(22-21) line is seen at 3.5\,$\sigma$ levels.  Based on our RADEX modeling, the SLED of the three lowest transitions is best fitted by gas with a kinetic 
temperature of 25\,K (dashed line). Examples of RADEX models that are consistent with the \her\ data are also shown for gas with a kinetic temperature of 400\,K, for 
a couple of column and volume density combinations (dashed-dotted and dotted lines).
The SLEDs of other prototype sources are overplotted for comparison, normalized to L$_{\rm {CO(1-0)}}$(4C12.50)/L$_{\rm {CO(1-0)}}$.
}
\label{fig:sled}
\end{center}
\end{figure*}

The IRAM 30m telescope \coto\ data (Figure~\ref{fig:CO21}) have a beam of $\sim$12\arcsec , which comprises all regions seen in emission in the PdB data. 
The measured \coto\ flux\footnote{
The width of the \coto\ line in the 30\,m telescope data, 390$\pm$68 \kms , is lower and more uncertain than that of the \cooz\ line in the PdB data. The uncertainty 
is due to the sinusoidal baseline in the single-dish data due to standing waves related to the bright continuum of 4C12.50. It does not necessarily translate into 
a flux loss. For comparison see the different line shape but identical line flux of \cooz\ obtained with the PdB array and the 30\,m telescope \citep{dasyra_combes12}.}
was 35 ($\pm$)6 Jy \kms , which together with a previously measured \cott\ flux of 50$\pm$8 Jy \kms\ \citep{dasyra_combes12} indicate a subthermal excitation 
of the CO molecules. Their excitation up to $J$=3 is identical with that of the gas in the Milky Way,  towards the Galactic center, and higher than that in outflow of M82 
\citep[Figure~\ref{fig:sled};][]{weiss01,weiss05}.

\begin{table}[hb!]
\begin{center}
\caption{CO Line Observations
\label{tab:codata}}
\scalebox{0.9}{
\begin{tabular}{llll}
\hline\hline
Line	   & $\lambda_{\rm restframe}$  & flux   & FWHM                    \\
  & (\micron )  & ($10^{-18}$ W\,m$^{-2}$)  &   (\kms )        \\
\hline
CO(1-0) & 2600.7 &  0.064$\pm$0.005 & 517$\pm$14  \\
CO(2-1) & 1300.4 &  0.24$\pm$0.04 &  390$\pm$68  \\
CO(3-2) & 866.96 &  0.51$\pm$0.08\tablefootmark{(1)} &  510$\pm$65\tablefootmark{(1)}   \\
CO(5-4) & 520.23 &  $<$8.0\tablefootmark{(2)}  & ...  \\
CO(6-5) &  433.55 &  $<$4.7\tablefootmark{(2)}  & ...  \\
CO(7-6) &  371.65 &  $<$3.5\tablefootmark{(2)}  & ...  \\
CO(8-7) &  325.22 &  $<$4.4\tablefootmark{(2)}  & ...  \\
CO(9-8) &  289.12 &  $<$5.8\tablefootmark{(2)}  & ... \\
CO(10-9) &  260.24 &  $<$6.0\tablefootmark{(2)}  & ...  \\
CO(11-10) &  236.61 & $<$6.0\tablefootmark{(2)}   & ...  \\
CO(12-11) &  216.93 & $<$6.0\tablefootmark{(2)}   & ...  \\
CO(13-12) &  200.27 & $<$6.0\tablefootmark{(2)}   & ...  \\
CO(16-15)  & 162.81 & $<$2.5  & ...  \\
CO(18-17)  & 144.78 & $<$1.8  & ... \\
CO(22-21)  & 118.58 & 0.97$\pm$0.27 &...      \\
CO(33-32)  & 79.36   &  $< $5.0 & ... \\
\hline
\end{tabular}
}
\end{center}
{\raggedright
\tablefoottext{1}{Value taken from \citet{dasyra_combes12}}.\\
\tablefoottext{2}{Estimated from SPIRE's sensitivity for 2.8\,hrs on source.}
}
\end{table}

\begin{table}[hb!]
\begin{center}
\caption{Continuum fluxes in the infrared and mm range}
\label{tab:contdata}
\begin{tabular}{ll}
\hline\hline
$\lambda_{\rm restframe}$  & flux          \\
(\micron )  & (Jy)    \\
\hline
52.0   & 2.21$\pm$0.10 \\
57.5   & 2.39$\pm$0.15 \\
78.7   & 2.46$\pm$0.11 \\
81.0   & 2.30$\pm$0.18 \\ 
121.0 & 1.51$\pm$0.06 \\
126.0 & 1.53$\pm$0.10 \\
146.0 & 1.48$\pm$0.12 \\
155.0 & 1.36$\pm$0.17 \\
158.5 & 1.15$\pm$0.10 \\
162.5 & 0.85$\pm$0.13 \\
172.0 & 1.00$\pm$0.19 \\
870 & 0.15$\pm$0.02 \\
1305 & 0.30$\pm$0.07 \\
2610 & 0.51$\pm$0.06 \\
\hline
\end{tabular}
\end{center}
\end{table}

\clearpage


\begin{figure}[h!]
\begin{center}
\includegraphics[width=9cm]{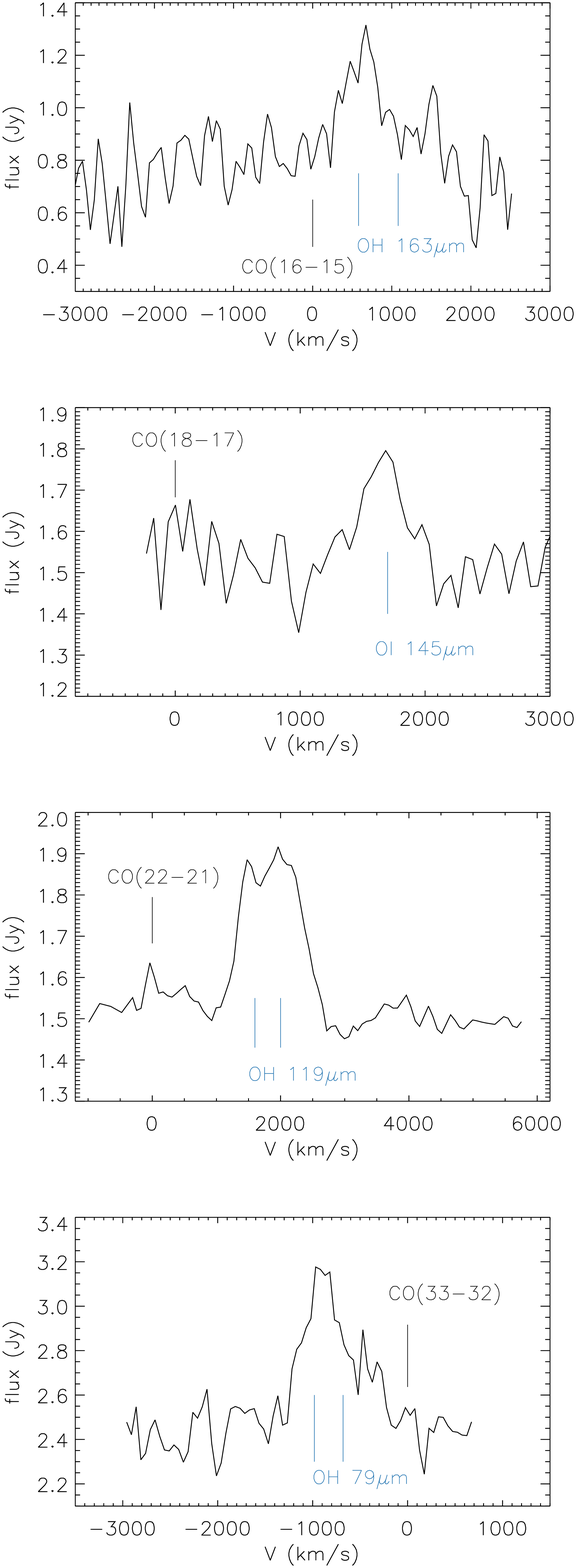} 
\caption{PACS observations of 4C12.50 that cover the wavelengths of high-$J$ CO spectral lines.}
\label{fig:pacsCO}
\end{center}
\end{figure}
\clearpage

\begin{figure*}
\begin{center}
\includegraphics[width=16.3cm]{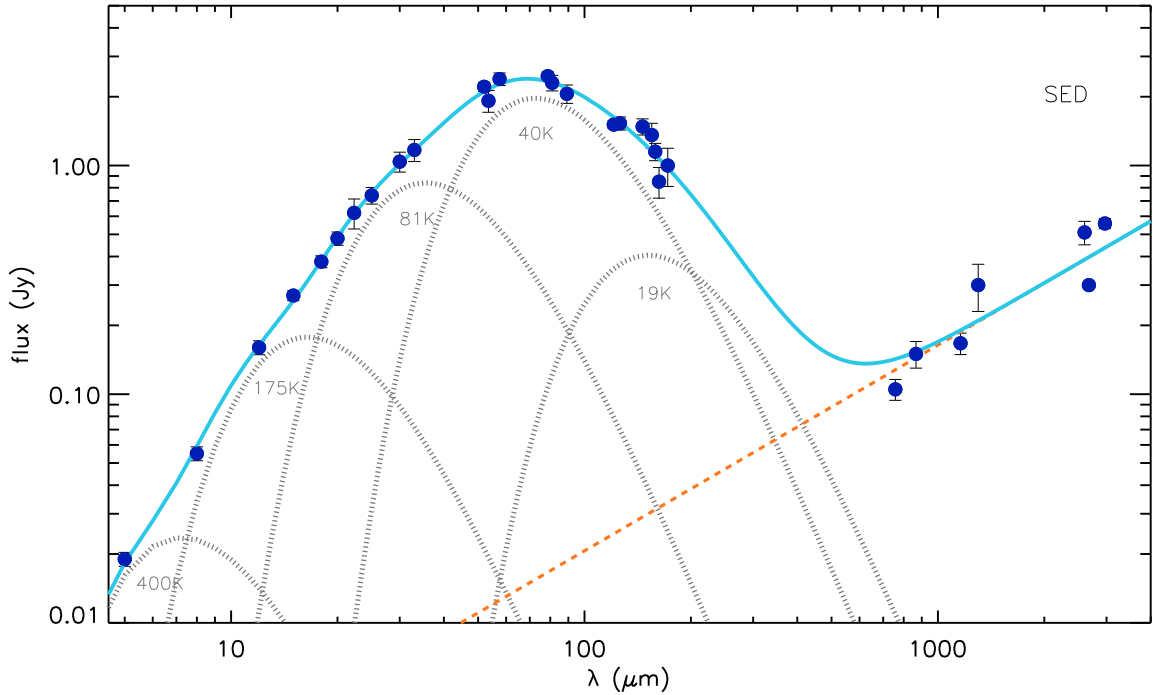} 
\caption{ Dust SED of 4C12.50 including continuum measurements from the new IRAM and \her\ data. The modified black body curves (with temperatures between 
19\,K and 400\,K) that best fit the IR/sub-mm data are plotted with dotted lines. The dashed line is a synchrotron-related power law with an exponent of 0.9. 
The solid line corresponds to the sum of the flux of all components.}
\label{fig:sed}
\end{center}
\end{figure*}

High-$J$ CO emission was not detected with PACS, with a probable exception for the 22-21 transition (Figure~\ref{fig:pacsCO}): weak (3.5$\sigma$) signal was seen at 
restframe 118.58\um , which was also seen by \citet{spoon13} and \citet{veilleux13} in their independent reductions of different sub-sets of the data. For the sake 
of completeness, we note that a spectrum of 4C12.50 showing no detection of CO lines was also acquired (OT1\_pogle01\_1) with SPIRE \citep{griffin10}.  
The limits used in this work (Table~\ref{tab:codata}) are based on the wavelength-dependent instrumental sensitivity for the integration time of the SPIRE observations. 

By means of a qualitative comparison of the SLED (Figure~\ref{fig:sled}), we find that the high-$J$ states of CO molecules are considerably less populated in 4C12.50 than 
those in local prototype AGN in which shock fronts and X-ray-photodissociation regions are likely to contribute to the gas excitation \citep[e.g., NGC1068 and NGC6240;][]{krips11,haileydunsheath12,spinoglio12,meijerink13}.

\subsection{CO spectral line energy distribution fitting: cold vs. warm gas reservoir}

To derive fundamental properties of the gas that is probed by our CO observations, we modeled the CO SLED with the radiative transfer code RADEX, assuming for 
simplicity that the flux that we detected over a line width of $\sim$520\kms\ originates from a collection of clouds with same average properties and intrinsic line widths 
of 1\kms\ \citep{vandertak07}. Further assuming that the cosmic microwave background temperature is 3.06\,K and that the CO is excited by collisions with hydrogen 
and helium, we ran a grid of CO column density $N_{CO}$, CO kinetic temperature \tkin , and \htwo\ volume density $n_{H2}$ models to find those that best fit our data. 

We considered intrinsic column densities in the range 5$\times$10$^{14}$$-$5$\times$10$^{17}$\,cm$^{-2}$,
corresponding to the beam-averaged CO column density in the chosen velocity step, divided by a filling factor of the beam by the clouds in the range 1-0.001. To compute this
beam-averaged column density from the \cooz\ observations, we used an \htwo\ gas mass of 1.0$\times$$10^{10}$\msun , a CO abundance that is 10$^4$ times lower than
that of the \htwo, and we equally divided the mass between 520 velocity bins. Having three unknown parameters and three line luminosity measurements, we obtained 
an exact solution for the model that best describes the molecular emission from the  $J$=1, $J$=2, and $J$=3 upper states. It corresponds to a filling factor of 0.025, a CO 
kinetic temperature of 25\,K, and an \htwo\ volume density of 1.5$\times$10$^{3}$\,cm$^{-3}$. In this solution, the \cooz -emitting gas is out of local thermodynamic 
equilibrium (LTE): the \cooz\ beam-corrected brightness temperature is 11\,K, i.e., $\sim$two times lower than the \htwo\ kinetic temperature. The data of the \cooz\ and \coto\ 
lines alone are compatible with a temperature of 7\,K, for the same volume density but for a higher filling factor (0.092). The data of the \cooz\ and \cott\ lines alone are compatible 
with a temperature as high as 32\,K, again for the same volume density but for a lower filling factor (0.017). 

To estimate the maximum warm gas mass that is probed by the CO, we calculated the maximum emission by molecules at high rotational states that is compatible 
with the observed SLED after fixing \tkin\ to 400\,K. This kinetic temperature is marginally higher than the excitation temperature of the \htwo\ gas seen with \spi\ 
\citep[374$\pm${12}\,K;][]{dasyra_combes11}. For this computation, our free parameters were the \htwo\ volume density and the CO column density. Our column
density grids were expanded to lower values than before.  A model fitting the CO(18-17) and CO(22-21) data has an \htwo\ volume density of 2$\times$10$^6$\,cm$^{-3}$,
and a CO column density of 5$\times$10$^{13}$ cm$^{-2}$ (Figure~\ref{fig:sled}). The mass of the warm \htwo\ as probed by the CO at 400\,K is then equal to 
2.5$\times$1$0{^7}$\msun , or 0.0025 times the mass of the cold \htwo\ probed by the CO at 25\,K, under the assumption that the emitting areas of the warm and
the cold gas are identical. The 400\,K gas mass would decrease if the CO(22-21) line luminosity were to be treated as an upper limit. If the warm gas were 
to be more tenuous, with an \htwo\ volume density of 1$\times$10$^5$\,cm$^{-3}$, then the CO column density would have to be below 5$\times$10$^{14}$ cm$^{-2}$ 
for the CO(7-6) to CO(13-12) luminosity limits to be respected. In that case, the upper limit on the mass of the warm gas would be 2.5$\times$10${^8}$\msun . In the 
likely presence of intermediate \tkin\ components, the mass estimate will again decrease.

To compare the gas and dust properties in the ambient medium, we produced the dust spectral energy distribution (SED) of 4C12.50 using our \her\ and IRAM continuum
flux measurements \citep[from this work, Table~\ref{tab:contdata}, ][]{dasyra_combes12} together with data from the literature \citep{moshir90,steppe95,clements10,
ostorero10,trippe10,guillard12}. The large scatter in the radio data is due to the time variability of the continuum. We fitted the SED by minimizing the $\chi^2$ between the 
observations and a model consisting of multiple modified black body curves and a power law. We parameterized the flux emitted by each modified black body as
\hbox{ $( M_{\rm dust}\, \kappa_{\rm abs}[\lambda_0] / D_L^2)\, (\lambda_0/\lambda )^{\beta}\,(2 h \nu ^3 / c^2)\, /(e^{h\nu /kT} -1)$ }, where $M_{\rm dust}$ is the dust mass 
at temperature $T$, $\kappa_{\rm abs}[\lambda_0]$ is the dust absorption coefficient at the reference wavelength $\lambda_0$, $h$ is the Planck constant, and $k$ is the 
Boltzmann constant. We opted for a reference wavelength accessible to PACS, 150\micron , and parameter values of $\beta$=2.0 and $\kappa_{\rm abs}$[150\micron]=11.6 cm$^2$\,gr$^{-1}$. 
These parameter values reproduce the absorption coefficient of the \citet{draine03} $R_V$=3.1, Milky Way dust model with an accuracy of 10\% in the 70 to 1000$\micron$ 
range. We limited the temperature of the coldest dust SED component in the range 5-35\,K. We also fixed the temperature of another component at 400\,K, i.e., at the \tkin\ 
of the warm CO SLED  component. A $\chi ^2$ minimization using the IDL MPFIT code indicated that three more components are needed to reproduce the 
observed SED (Figure~\ref{fig:sed}). From the SED modeling, we found that the infrared luminosity of 4C12.50 is 2.4($\pm$0.1)$\times$10$^{12}$\lsun , that
the component that best fits the peak of the SED is at $T$=40$\pm$3\,K, and that the component with the lowest temperature is at 19\,$\pm$5K.  

The temperatures of the coldest dust SED components thus bracket the kinetic temperature of the cold CO SLED component (25$\pm$8\,K). The SLED could 
also be reproduced up to CO(3-2) by a combination of two gas components with temperatures lower than or equal to those of the dust (e.g., 7\,K and 40\,K). Collisions 
with the ambient ISM could thus suffice to explain the excitation of the bulk of the CO molecules. A jet-induced shock does not have to be invoked instead. 


\section{Discussion}
\label{sec:discussion}

\subsection{Heating of the accelerated molecular gas with respect to the ambient medium}

The results presented in Section~\ref{sec:results} raise the question of how do the excitation properties of the accelerated molecular gas
compare with those of the dynamically relaxed gas. From the \spi\ data, we know that the outflow has a warm component, carrying 5.2$\times$10$^7$\msun\ 
of  \htwo\ molecules at $\sim$400\,K \citep{dasyra_combes11}. From a CO(2-3) absorption line that is due to clouds obscuring the millimeter continuum 
and moving at -950($\pm$90)\kms , we know that colder molecules do exist in the outflow \citep{dasyra_combes12}. Only a lower limit could be placed 
on their mass from the CO(2-3) data, because the absorption only probes the few lines of sight towards the millimeter continuum. The millimeter continuum 
is less extended than the radio continuum due to the steep power law that describes the emission from the jet lobes \citep{krichbaum98, krichbaum01, 
krichbaum08}, and the radio continuum is on its turn constrained to $\sim$ten hot spots with a radius of a few parsec each \citep{morganti13b}. The bulk 
of the cold entrained CO is thus likely to be found in lines of sight without a millimeter background, and it should be detected in emission. Its non-detection 
in our \cooz\ PdB data now enables us to place an upper limit of 1.3$ \times$10$^{8}$\msun\ on the mass of cold entrained \htwo\ gas. 

Putting all numbers together, we find that the mass ratio of warm (400\,K) to cold (25\,K) \htwo\ gas in the outflow is $>$0.40. For the ambient medium, containing 
1.0$\times$10$^{10}$\msun\ of \htwo\ molecules at 25\,K and 1.4$\times$10$^8$\msun\ of \htwo\ molecules at $\sim$400\,K \citep[Section~\ref{sec:results};][]{dasyra_combes11}, 
the same ratio is only 0.014. A major result of this work is thus that the mass ratio of warm-to-cold \htwo\ gas is at least 29 times higher in the outflow than in the 
ambient medium. 

\begin{figure}
\begin{center}
\includegraphics[width=\columnwidth]{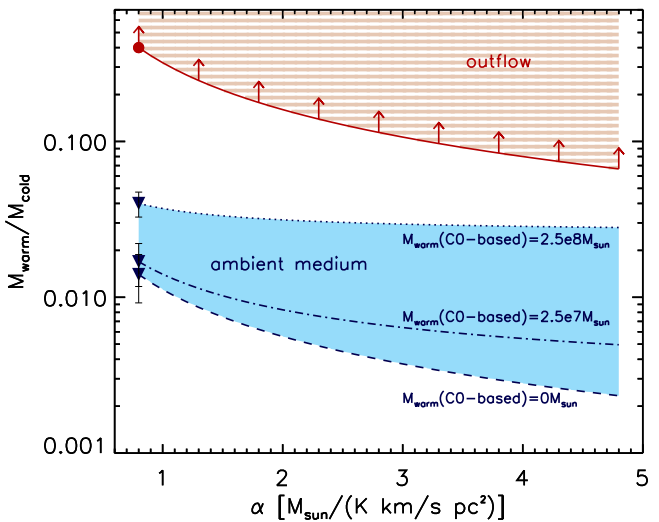} 
\caption{Warm-to-cold gas mass ratio as a function of the $\alpha$ conversion factor. The circle shows the ratio's lower limit in the outflow for  
$\alpha$=0.8\msun /(K\kms\,pc$^2$). The solid line shows how this lower limit decreases as $\alpha$ approaches values thought appropriate for
 the Milky Way. This curve brackets the range of all acceptable M$_{warm}$/M$_{cold}$ values in the outflow (hatched area), under the assumption that the mass 
 of the 400\,K outflowing gas probed by the CO is negligible with respect to that directly probed by the \htwo. For the ambient medium,  the measured ratio (triangles) 
 and its dependence on $\alpha$ (non-solid lines) are given for different warm gas content scenarios:  when only the mass of the 400\,K \htwo\ gas directly 
 seen by \spi\ is used (dashed line), and when the mass of another CO-based 400\,K gas component is added. The dashed-dotted line uses an additional 
 mass of 2.5$\times$10${^7}$\msun\ distributed in clouds of 2$\times$10$^6$\,cm$^{-3}$ (as in the dashed-dotted  CO SLED component in Figure~\ref{fig:sled}). 
 The dotted line instead uses an additional mass of 2.5$\times$10${^8}$\msun\ distributed in clouds of 1$\times$10$^5$\,cm$^{-3}$ (as in the dotted CO SLED 
 component in Figure~\ref{fig:sled}).  This line constrains the upper range of acceptable M$_{warm}$/M$_{cold}$ values in the ambient medium (solid area).} 
\label{fig:massratio}
\end{center}
\end{figure}

Taking into account that a non-negligible fraction of the mid-infrared \htwo\ emission could be due to gas that was accelerated by the front shock but that is not accounted 
for in the outflow due to  its low velocity, we conclude that the discrepancy in the warm-to-cold mass ratio could increase. Contrarily, adding the mass of the warm ambient 
\htwo\ probed by high-$J$ CO lines to the mass of the warm ambient \htwo\ measured by \spi\ could decrease this discrepancy. Still, it would not alter our conclusion. 
Adding 2.5$\times$10${^7}$\msun\ of gas distributed in a dense 400\,K component (in agreement with the high-$J$ CO emission in Figure~\ref{fig:sled}), would lead to 
a warm-to-cold \htwo\ mass ratio that is $\ge$24 times higher in the outflow than in the ambient medium. Even adding the maximum amount of 400\,K gas that is 
compatible with the CO SLED, 2.5$\times$10${^8}$\msun , would lead to a ratio that is $\ge$10 times higher in the outflow than in the ambient medium. Likewise, 
our conclusion is robust against the use of another $\alpha$ factor, or the use of different $\alpha$ factors for the entrained and ambient gas \citep[see, e.g.,][]{papadopoulos12}. 
The latter choice can be justified because the CO intensity to \htwo\ mass conversion scales with \tkin $^{-1}$$(n_{H2})^{1/2}$ for virialized clouds \citep{maloney88,weiss01}. 
Figure~\ref{fig:massratio}, showing the dependence of the warm-to-cold \htwo\ mass ratio on the $\alpha$ factor, indicates that there is no combination of acceptable 
$\alpha$ and warm gas mass values that could make the ratio in the ambient medium agree with that in the outflow. Instead, the ratio in the outflow is at least 3$-$30 
times higher than that in the ambient medium in the entire parameter space.

Our results constitute direct evidence that the accelerated molecular gas can be heated\footnote{
Heating of the molecular gas, leading to excitation at high rotational states, could be one of the reasons why absorption of the mm continuum by outflowing molecules 
is not (reliably) detected in CO(0-1) in the PdB data while it was detected in CO(2-3) in the single-dish data. Alternatively, absorption in CO(0-1) may not be seen if it is 
diluted in CO(1-0) emission. A similar scenario may not hold for CO(2-3) and CO(3-2) depending, e.g., on the size and the geometry of the emitting regions. Changes with time 
in the relative position of the foreground clouds and the background source leading to changes in the covering factor, or changes with time in the millimeter continuum flux 
could also justify this result.} 
to the point of being inefficient to form new stars, and that it can be detected 
even though the cooling timescales can be considerably shorter than the outflow propagation timescales \citep{guillard12}, possibly thanks to the enhancement of 
turbulence on individual-cloud scales \citep{nesvadba11}. \citet{papadopoulos08} and \citet{ogle10} previously attributed the overall strong emission by CO molecules 
in states of intermediate rotational number and by \htwo\ molecules in states of low rotational number to shock excitation by radio jets. 

\subsection{Impact of the outflow on the galaxy}
\label{sec:impact}

The outflow could affect the future evolution of 4C12.50 via gas heating or gas expulsion. While the effects of heating are uncertain, we observe that the feedback mechanism has not had 
the time to reach and excite a large fraction of the CO reservoir.

To test the role of gas expulsion, we compute the mass outflow rate, $\dot{M}_{out}$, which is equal to \hbox{$M_{out} V_{out} d^{-1}$}. We revise the flow rate of warm gas
presented in \citet{dasyra_combes11} from 130\msun yr$^{-1}$ to 230\msun yr$^{-1}$, using identical values for the outflow velocity $V_{out}$ (640\kms ) and mass $M_{out}$ 
(5.2$\times$10$^7$\msun ), but assuming that the entrained gas is within a distance $d$ of 150\,pc from the radio core \citep{morganti13b}. The flow of cold gas could carry up 
to 570\msun yr$^{-1}$, under the assumption that the entrained CO and the entrained \htwo\ have the same average velocity. This assumption is not in conflict with the CO(2-3) detection 
at $-$950\kms\ \citep{dasyra_combes12}, which only probes the motions of some outflowing clouds: those situated between us and the radio core. For both gas phases, the total mass flow 
rate is in the range 230-800\msun yr$^{-1}$.  

However, only a part of this gas could be lost to the intergalactic medium. The escape velocity at 150\,pc from the west nucleus of 4C12.50 is 700\kms , when assuming an isothermal sphere 
distribution for the visible baryonic matter and when using a stellar velocity dispersion of 170\kms\ and a half-light radius of 2.6\,kpc \citep{dasyra06a}. The escape velocity will increase to 
800\kms\ when the mass in the east nucleus of 4C12.50 is accounted for.  From the \htwo\ velocity distribution in the \spi\ data, less than 30\% of the gas could be lost to the 
intergalactic medium. Most of the accelerated and heated gas will fountain back to a disk within a few dynamical timescales. The maximum mass loss rate would be $\lesssim$240\msun /yr, 
and the minimum depletion timescale of the reservoir $\gtrsim$4$\times$10$^7$ yrs. The depletion timescale could increase further and exceed the typical values for AGN duty cycles when 
the mass in the dark matter halo(s) is taken into account. 

Any gas expulsion could be moderated or counteracted by a potential inflow of gas from the southern tidal tail. At the projected distance of the tail, $\sim$10\,kpc, the escape velocity from 
the visible baryonic matter is of order 200-300\kms , i.e., higher than the typical rotational velocities of spiral arms. Given that the line-of-sight velocity of the tidal tail is $<$40\kms , gas could be 
captured and fed to the western nucleus.

We thus postulate that both gas expulsion and heating are likely to temporarily delay but unlikely to entirely suppress star formation in 4C12.50.

\subsection{Driving mechanism of the outflow}

Power and momentum arguments are typically invoked for the evaluation of the outflow driver. The flow kinetic luminosity can be computed from the product 
$0.5 M_{out} V_{out}^3 d^{-1}$. For the $M_{out}$, $V_{out}$, and $d$ values discussed in Section~\ref{sec:impact}, the kinetic luminosity of the warm entrained gas that is probed 
by the \htwo\ is 3$\times$10$^{43}$\,erg\,s$^{-1}$. Adding the maximum amount of cold outflowing gas that could be probed by the CO would place the kinetic luminosity upper limit to 
1$\times$10$^{44}$\,erg\,s$^{-1}$. The luminosity radiated by the outflowing gas in the purely rotational \htwo\ lines, 4$\times$10$^{41}$\,erg\,s$^{-1}$  \citep{dasyra_combes11}, is 
negligible with respect to the outflow kinetic luminosity. The luminosity radiated in CO lines is even lower than that radiated in \htwo\ lines.  
 
According to \citet{veilleux09}, half of the infrared luminosity of this source can be ascribed to its AGN and half to its starburst, with an uncertainty of $\sim$50\% on either fraction. The 
dust-based star-formation-rate estimate is then 400\msun /yr \citep{kennicutt98}. The power released by supernovae (SN) is 10$^{44}$ erg\,s$^{-1}$, under the assumption that  for every 
hundred solar masses forming there is one SN \citep[e.g., in the][initial mass function context]{kroupa01}, ejecting material with a kinetic energy of 10$^{51}$\,erg. This energy is insufficient 
to drive the flow unless a very large fraction (30-100\%) of it is converted into the outflow kinetic luminosity.

The luminosity of the AGN, like that of the starburst, is estimated to be 5$\times$10$^{45}$\,erg\,s$^{-1}$.  The force exerted to the gas by radiation pressure is then 5-18 times lower 
than the outflow momentum rate, when the latter is approximated by the product $\dot{M}_{out}$$V_{out}$ \citep{combes13}. The momentum rate boost needed for the gas acceleration 
could occur with the aid of energy-conserving gas expansion or multiple photon scatterings in high optical depth regions \citep{faucher12}. For these mechanisms, the compact-source
geometry of the AGN is advantageous.

The AGN jet constitutes a generous power source, capable of driving the flow. The radio cavity kinetic luminosity is estimated from its 178\,MHz flux to be 3$\times$10$^{45}$\,erg\,s$^{-1}$ 
\citep{guillard12}. The most significant argument in favor of the radio jet remains the location of the H\,I outflow, at the tip of the southern radio hot spot \citep{morganti13b}. The young age of the jet,  
limited to $<$10$^5$ yrs \citep{lister03}, could explain why only a small amount of the CO is affected by the feedback. 

\subsection{Implications for future observations}

Recent hydrodynamic simulations indicate that gas heating could be common in AGN outflows: the effects of jets and radiation-driven ultra-fast winds on the ISM can be comparable 
only 10$^4$-10$^5$ years after the onset of the feedback activity \citep{wagner11,wagner13}, when both mechanisms drive bubbles of similar flow structure and dynamics.

If heating of the accelerated molecular gas is common, then the detection of outflows could be easier in intermediate-$J$ CO lines than in low-$J$ CO lines, facilitating the discovery of 
outflows in the distant Universe: at z$\sim$1 or higher, for example, ALMA could probe outflows of comparable mass to those presently detected at 100\,Mpc with the PdB. Moreover, the acquisition 
of infrared data alongside millimeter data could be valuable for the systematic discovery and reliable mass determination of outflows. With ALMA operating in full-array mode in $\sim$two 
years and with the {\it James Webb} Space Telescope coming up beyond 2018, these goals will be achievable for most systems of interest in the local Universe. At 100 Mpc, ALMA is 
designed to detect 5$\times$10$^5$\msun\ outflows of 10\,K gas in \cooz . The JWST is designed to reach the same limit for the $\sim$100K gas probed by \htwo (0-0)S(1). For typical outflow 
radial extents of a few hundred pc, this common detection threshold corresponds to an outflow rate of $\sim$1\msun yr$^{-1}$, comparable to the star formation rate in local spirals.

\section{Conclusions}
\label{sec:conclusions}
 
We used IRAM PdB, 30\,m telescope, and \her\ Space Telescope data to study the  content, excitation, and kinematic properties of the molecular gas in the outflow and in the ambient 
medium of a local prototype radio-loud, ultraluminous-infrared galaxy, 4C12.50  (also known as IRAS13451$+$1232). We specifically looked for gas heating and expulsion signatures, 
i.e., characteristic of the two feedback-related processes that can delay star formation. Combining new and previous observations, we found evidence for both processes taking place,
and we drew the following main conclusions.

\begin{itemize}
\item
The \cooz\ emission spatially consists of a newly discovered tidal tail, which extends out to 12\,kpc south-west of the nucleus, and a main component, which is marginally resolved
within the beam radius (4.2 kpc). A shift in the location of the maximum emission with velocity indicates that the main component could be tracing either two merging nuclei or a 
rotating disk. Its reservoir contains 1.0($\pm$0.1)$\times$10$^{10}$\msun\ of cold \htwo\ gas for $\alpha$=0.8\msun /(K\kms\,pc$^2$).
\item
The CO SLED, which is primarily tracing the ambient gas in 4C12.50, is very similar with that of the Milky Way gas for low-$J$ numbers. Fitting of the CO SLED with RADEX indicates 
that the gas mass probed by CO molecules at 400\,K is less than 0.025 times the gas mass probed by CO molecules at 25\,K. Likewise, the \htwo\ gas mass probed by CO molecules 
at 400\,K is less than 1.7 times the mass of the 400\,K gas probed by the purely rotational \htwo\ lines in the mid-infrared.
\item 
In the outflow, the mass of the cold (25\,K) \htwo\ gas is $<$1.3($\pm$0.1)$\times$10$^8$\msun , i.e., at most twice as high as the mass of the warm (400\,K) \htwo\ gas emitting in 
the mid-infrared. By combining the \htwo -based mass flow rate measurement from the mid-infrared data, 230\msun /yr, with the CO-based mass flow rate upper limit from the mm 
data, 570\msun /yr, we conclude that the feedback mechanism(s) could be accelerating up to 800\msun /yr of molecular gas for an outflow radius of 150\,pc.  
\item
A low fraction of the outflowing gas ($<$30\% when ignoring dark matter) could escape the system and be lost to the intergalactic medium. Any mass loss could be moderated or 
counteracted by a potential inflow of gas from the tidal tail containing 3.3($\pm$0.3)$\times$10$^8$\msun\ of cold \htwo . The star formation in the host of 4C12.50 is likely 
to be delayed, but not entirely suppressed by the feedback.
\item
The mass ratio of warm-to-cold \htwo\ gas is elevated in the outflow with respect to the ambient medium by a factor of $\gtrsim$30 for $\alpha$=0.8\msun /(K\kms\,pc$^2$).
The conclusion that the accelerated molecular gas is heated is robust against two main sources of uncertainty: the chosen $\alpha$ conversion factor, and the mass of the warm 
gas probed by the CO. While the discrepancy will be lower for higher values of either parameter, there is no combination of acceptable parameter values that could make the 
warm-to-cold gas mass ratio in the ambient medium reach that in the outflow.
\item
A scenario in which outflows could be probed by a different CO SLED than the rest of the ISM is likely. It remains to be tested with sub-millimeter observations.
\end{itemize}


\begin{acknowledgements}
K. M. D.  thanks  F. Casoli, O. La Marle, and M. Lozach for making this project possible by means of a French Space Agency (Centre National d'\,Etudes Spatiales; CNES) fellowship. 
Our work is based on data obtained with the facilities of IRAM, which is supported by INSU/CNRS (France), MPG (Germany), and IGN (Spain), and on data obtained with the \her\ Space Telescope, 
which is an ESA space observatory with science instruments provided by European-led principal investigator consortia and with important participation from NASA.
\end{acknowledgements}

{}

\end{document}